# *Ab initio* determination of effective electron-phonon coupling factor in copper

Pengfei Ji and Yuwen Zhang[1]

Department of Mechanical and Aerospace Engineering

University of Missouri

Columbia, MO 65211, USA

## Abstract

The electron temperature $T_e$ dependent electron density of states $g(\varepsilon)$, Fermi-Dirac distribution $f(\varepsilon)$, and electron-phonon spectral function $\alpha^2 F(\Omega)$ are computed as prerequisites before achieving effective electron-phonon coupling factor $G_{e-ph}$. The obtained $G_{e-ph}$ is implemented into a molecular dynamics (MD) and two-temperature model (TTM) coupled simulation of femtosecond laser heating. By monitoring temperature evolutions of electron and lattice subsystems, the result utilizing $G_{e-ph}$ from *ab initio* calculation, shows a faster decrease of $T_e$ and increase of $T_l$ than those using $G_{e-ph}$ from phenomenological treatment. The approach of calculating $G_{e-ph}$ and its implementation into MD-TTM simulation is applicable to other metals.

## 1. Introduction

When a metallic system is under the conditions of neutron irradiation, swift heat ion and gamma radiation, as well as femtosecond laser heating, the electron temperature $T_e$, increases to tens of thousands of degrees Kelvin as a result of electron excitation. At a low lattice temperature $T_l$, the lattice subsystem remains unaffected. The large increase of $T_e$ and relative cold $T_l$ results in electron-phonon non-equilibrium. This strong non-equilibrium between the electron subsystem and lattice subsystem of a metal affects a large number of physical properties, such as the thermal conductivity, superconductivity and thermal stress propagation.

During the transport of the absorbed thermal energy from electron subsystem to lattice subsystem, the electron-phonon coupling factor $G_{e-ph}$, plays a crucial role in influencing the relaxation time needed to reach equilibration state. In the past 40 years, a large body of research has been performed to study electron-phonon interaction and thereby determine $G_{e-ph}$. The electron-phonon coupling constant $\lambda_{const}$ was primarily proposed by Allen *et al.* [1, 2] to characterize the strength of electron-phonon coupling and was subsequently measured by using the pump-probe procedure [3]. Lin and Zhigilei proposed a $T_e$ dependent $G_{e-ph}$ and carried out related calculations by including the experimental $\lambda\langle\omega^2\rangle_{const}$ [4]. A phenomenological approach to calculating temperature dependent $G_{e-ph}$ was derived by Chen *et al.* [5], which included both the effects of electron-electron and electron-phonon scattering. Many *ab initio* calculations were also performed to study $G_{e-ph}$ [6–8]. Nevertheless, there is still no work that comprehensively treats all $T_e$ dependent parameters (Fermi-Dirac distribution $f(\varepsilon)$, electron density of states $g(\varepsilon)$ and electron-phonon spectral function $\alpha^2 F(\Omega)$) in obtaining $G_{e-ph}$.

---

[1] Corresponding author. Email: zhangyu@missouri.edu

This letter paves a new way of pure *ab initio* calculation of $G_{e-ph}$ after derivation of all the essential parameters: $g(\varepsilon)$, $f(\varepsilon)$ and $\alpha^2 F(\Omega)$.

## 2. Computational details

Recalling the definition of $G_{e-ph}$, the effective electron-phonon coupling can be mathematically expressed by means of the rate of energy transportation $\partial E/\partial t$ per unit cell volume $V_c$ at the temperature difference between $T_e$ and $T_l$ [2]

$$G_{e-ph} = \frac{\partial E}{\partial t} \frac{1}{T_e - T_l} \frac{1}{V_c} \quad (1)$$

When electrons are excited, the variations of $g(\varepsilon)$, $f(\varepsilon)$ and $\alpha^2 F(\Omega)$ from the room temperature contribute to $\partial E/\partial t$ between electron and lattice subsystems. By taking the electron-phonon collision into account, $\partial E/\partial t$ can be obtained as [2]

$$\frac{\partial E}{\partial t} = \frac{4\pi}{\hbar} \sum_{k,k'} \hbar \omega_Q |M_{k,k'}|^2 S(k,k') \delta(\varepsilon_k - \varepsilon_{k'} + \hbar \omega_Q) \quad (2)$$

where $\hbar$ is the reduced Planck constant $1.054 \times 10^{-34}$. $k$ and $k'$ are the electron quantum number at initial and final states, respectively. $\omega_Q$ denotes the phonon frequency at the phonon quantum number $Q$. The scattering probabilities of electrons at initial energy $\varepsilon_k$ and final energy $\varepsilon_{k'}$ are described by the matrix $M_{k,k'}$. $S(k,k')$ equals $[(f_k - f_{k'})n_Q - f_{k'}(1 - f_k)]$, which is named as the thermal factor. $f_k$ is the Fermi distribution function for electron, $1/[\exp(\frac{\varepsilon - \mu}{k_B T_e}) + 1]$. $n_Q$ is the Bose distribution for phonon, $1/[\exp(\frac{\hbar \omega_Q}{k_B T_l}) - 1]$.

By introducing electron-phonon spectral function at a specified $T_e$

$$\alpha^2 F(\varepsilon, \varepsilon', \Omega)|_{T_e} = \frac{2}{\hbar g(\varepsilon_F)|_{T_e}} \sum_{k,k'} |M_{k,k'}||_{T_e}^2 \delta(\omega_Q - \Omega) \delta(\varepsilon_k - \varepsilon) \delta(\varepsilon_{k'} - \varepsilon') \quad (3)$$

and combining with Eq. (1) and (2), $G_{e-ph}$ at specified $T_e$ becomes

$$G_{e-ph}|_{T_e} = \frac{2\pi g(\varepsilon_F)|_{T_e}}{T_e - T_l} \frac{1}{V_c} \{\int_0^\infty [\int_{-\infty}^\infty (\int_{-\infty}^\infty \alpha^2 F(\varepsilon, \varepsilon', \Omega)|_{T_e} d\varepsilon) d\varepsilon'] \hbar \Omega d\Omega\} S(\varepsilon, \varepsilon')|_{T_e} \delta(\varepsilon - \varepsilon' + \hbar\Omega) \quad (4)$$

The energy conservation requires that $\varepsilon' - \varepsilon = \hbar\Omega$, and the electron-phonon spectral function is approximated as [9]

$$\alpha^2 F(\varepsilon, \varepsilon', \Omega)|_{T_e} = g(\varepsilon)|_{T_e} g(\varepsilon + \hbar\Omega)|_{T_e} \alpha^2 F(\varepsilon_F, \varepsilon_F, \Omega)|_{T_e} / [g(\varepsilon_F)|_{T_e}]^2 \quad (5)$$

Moreover, at the limit of $k_B T_e \gg \hbar\Omega$ and $k_B T_l \gg \hbar\Omega$, the thermal factor becomes

$$S(\varepsilon, \varepsilon')|_{T_e} = [f(\varepsilon) - f(\varepsilon + \hbar\Omega)](T_e - T_l)k_B/(\hbar\Omega) \quad (6)$$

Because the energy range of electrons is much wider than that of phonons, $g(\varepsilon)$ is approximately equal to $g(\varepsilon + \hbar\Omega)$ and $[f(\varepsilon) - f(\varepsilon + \hbar\Omega)]/(\hbar\Omega)$ in Eq. (6), which can be rewritten as $-\partial f/\partial \varepsilon$. In addition, at the high $T_e$ limit, the second moment of $\alpha^2 F(\Omega)|_{T_e}$ is simplified as $\lambda\langle\omega^2\rangle|_{T_e} = 2\int_0^\infty \alpha^2 F(\Omega)|_{T_e} \Omega d\Omega$ [2] Therefore, Eq. (4) becomes

$$G_{e-ph}|_{T_e} = \frac{\pi \hbar k_B \lambda \langle \omega^2 \rangle|_{T_e}}{[g(\varepsilon_F)|_{T_e}]^2} \frac{1}{V_c} \int_{-\infty}^{\infty} [g(\varepsilon)|_{T_e}]^2 (-\frac{\partial f|_{T_e}}{\partial \varepsilon}) d\varepsilon \qquad (7)$$

This letter focuses on the effective electron-phonon coupling factor of copper. The *ab initio* investigation of electron-phonon interaction is carried out by using the density function theory (DFT) code ABINIT [10]. The finite temperature density functional formalism proposed by Mermin [11] is utilized to represent the different degrees of electron excitation at given $T_e$. The thermalized electrons obey Fermi-Dirac distribution and are expressed at Kohn-Sham eigenstates in each self-consistent field calculation. The calculations are based on the local density approximation (LDA) in combining with the projector-augmented wave (PAW) atomic data [12] to obtain $g(\varepsilon)$ (shown in Fig. 1(a)) and $f(\varepsilon)$ (shown in Fig. 1(b)), and the norm-conserving pseudopotential method in cooperation with a linear response approach [13] to get $\lambda \langle \omega^2 \rangle$ at given $T_e$ (shown in Fig. 2(c)). The linear response calculation of electron-phonon interaction is performed under the framework of the linear-muffin-tin-orbital (LMTO) method [14]. The valence electrons for copper are $3d^{10}4s^1$. After the convergence test, the Monkhorst-Pack $k$-points grid of $18 \times 18 \times 18$ and a cutoff energy of $32\ Ha$ are chosen.

## 3. Results and discussion

The zero point of the horizontal axis in Fig. 1 is set as the Fermi energy $\varepsilon_F$. As seen in Fig. 1(a), as a result of the thermal excitation of the electron subsystem, $g(\varepsilon)$ at increased $T_e$ presents the overall shift toward lower ε state. Figure 1(a) shows gradual shrinking of the $d$ block, and similar $d$ block changes were reported in [15,16]. The latter shrinkage is said to be due to a more attractive electron-ion potential as a result of a decrease in electronic screening of $3d$ block electrons. Additionally, with the left shift and shrinkage of $g(\varepsilon)$, its value slightly increases. The Fermi-Dirac distribution $f(\varepsilon)$ at $T_e$ of $300\ K$, $1 \times 10^4\ K$, $2 \times 10^4\ K$ and $4 \times 10^4\ K$ is shown in Fig. 1(b). When $T_e = 300\ K$, it can be seen that $f(\varepsilon)$ focuses around the Fermi energy. With the continuously increasing $T_e$, $f(\varepsilon)$ gets smeared out. Some of the electrons at high $T_e$ are thermally excited at ε above the Fermi energy $\varepsilon_F$. As seen in Fig. 1(b), when $f(\varepsilon) = 0.5$ at $T_e = 4 \times 10^4\ K$, ε locates at the right side energy point of all the other three cases when $f(\varepsilon) = 0.5$, which indicates the increase of chemical potential $\mu$ at high $T_e$.

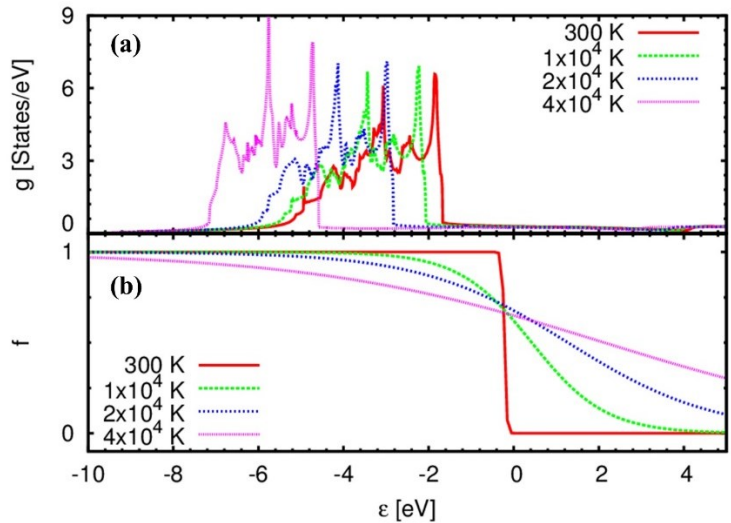

Fig. 1 (a) Electron density of states and (b) Fermi-Dirac distribution for electrons at $300\ K$, $10,000\ K$, $20,000\ K$ and $40,000\ K$.

In order to present more detailed information of $\mu$ at high $T_e$, the result of $\mu - \varepsilon_F$ is plotted in Fig. 2(a). It can be seen that $\mu$ increases with increasing $T_e$. The result in Fig. 2(a) agrees with that shown in [15], which was obtained from *ab initio* calculation by using generalized gradient approximation (GGA) for exchange and correlation energy. The computed $g(\varepsilon_F)$ is shown in Fig. 2(b), which presents a gradually decreasing tendency with the increment of $T_e$. Since $g(\varepsilon_F)$ is the dominator in the right hand of Eq. (7), the gradually decreasing $g(\varepsilon_F)$ brings the positive impact to $G_{e-ph}$, which indicates that $G_{e-ph}$ will be greater and greater under the continuous increment of $T_e$ (assuming all the other $T_e$ dependent parameters are kept as constants).

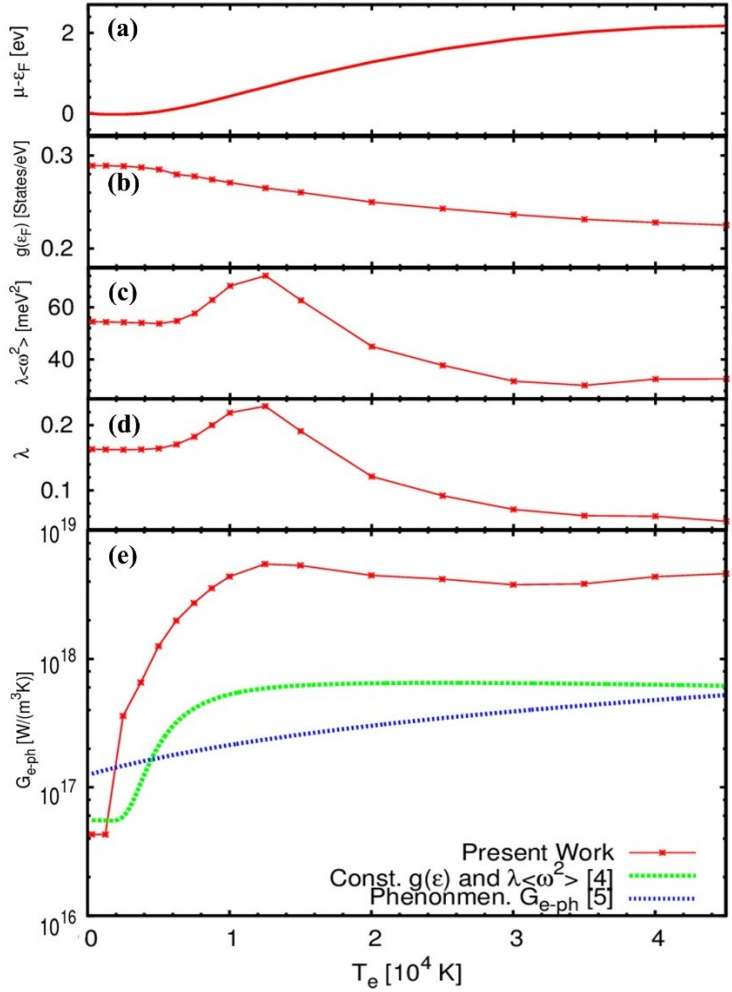

Fig. 2 Electron temperature $T_e$ dependent (a) relative chemical potential $\mu$ changes to Fermi energy $\varepsilon_F$, (b) electron density of states $g(\varepsilon)$ at the Fermi energy $\varepsilon_F$, (c) the second momentum of second moment of $\alpha^2 F(\Omega)$: $\lambda\langle\omega^2\rangle$, (d) the electron-phonon coupling factor $\lambda$, and (e) the effective electron-phonon coupling factor $G_{e-ph}$.

The electron-phonon spectral function $\alpha^2 F(\Omega)$ at given $T_e$ is calculated from response linear calculation. For the purpose of comparison, the experimental result [3] and *ab initio* calculation result [13] are plotted in Fig. 3 together with $T_e$ dependent $\alpha^2 F(\Omega)$ in the present work. The *ab* *initio* calculation result in [13] presents the highest peak when $\Omega$ is at $7\ THz$, which is not seen in the experimental result in [17] and all the results in present calculation in Fig. 3. It should be noted that $\alpha^2F(\Omega)$ calculated in [13] did not take the finite temperature density functional theory into account. However, in the practical application of laser heating and neutron irradiation, the electron subsystem excites after photon energy deposition. For the experimental study of $\alpha^2F(\Omega)$ by using the technique of forming tiny point contacts between copper and copper when electric voltage was added, the degree of electron excitation is unknown in the experiment [17]. Moreover, as seen in Fig. 3, frequency $\Omega$ locates from $0.8\ THz$ to $9.6\ THz$, which indicates that the electron-phonon relaxation time is around the picosecond scale. With the continuous thermal excitation of the electron subsystem, main peaks of $\alpha^2F(\Omega)$ (calculated in the present letter) shift toward the higher $\Omega$ region. Hence, it can be interpreted that the electron-phonon interaction becomes shorter at a higher degree of electron excitation. Accompanying the right shift of $\alpha^2F(\Omega)$, the electron-phonon interaction value decreases after $T_e = 1.0 \times 10^4\ K$; hence, a weaker electron-phonon interaction occurs at higher $T_e$.

On the basis of calculated $\alpha^2F(\Omega)$, $\lambda$ and $\lambda\langle\omega^2\rangle$ are computed to obtain $G_{e-ph}$ in Eq. (7). Figure 2(c) shows the results of $\lambda\langle\omega^2\rangle$ at given $T_e$. When $T_e < 0.8 \times 10^4\ K$, $\lambda\langle\omega^2\rangle$ is around $54\ meV^2$. When $T_e$ increases, $\lambda\langle\omega^2\rangle$ shows a peak (at $T_e = 1.4 \times 10^4\ K$) and decreases to $32\ meV^2$ when $T_e = 1.4 \times 10^4\ K$. Recalling $\lambda\langle\omega^2\rangle|_{T_e} = 2\int_0^\infty \alpha^2F(\Omega)|_{T_e}\Omega d\Omega$ and combining with Fig. 3, the peak of $\lambda\langle\omega^2\rangle|_{T_e=1.4\times10^4\ K}$ is mainly caused by slightly increasing $\alpha^2F(\Omega)$ and a right shift of $\alpha^2F(\Omega)$ toward high $\Omega$. As a non-dimensional parameter quantitatively assessing the strength of electron-phonon interaction without taking $\Omega$ into consideration, $\lambda$ is calculated and shown in Fig. 2(d). Almost similar profiles of $\lambda\langle\omega^2\rangle$ and $\lambda$ are achieved in Figs. 2(c) and 2(d). However, after $T_e = 3.5 \times 10^4\ K$, $\lambda$ shows lower value than that at $T_e = 3.5 \times 10^4\ K$. On the contrary, $\lambda\langle\omega^2\rangle$ shows the opposite trend induced by the right shift of $\alpha^2F(\Omega)$, which helps to compensate for the continuous decrement of $\lambda$. Beaulac *et al.* [18] calculated $\lambda$ as $0.116$ and $\lambda\langle\omega^2\rangle$ as $44.897\ meV^2$. The other *ab initio* linear response calculation result showed $\lambda$ as $0.14$ [13]. Brorson *et al.* [3] experimentally measured $\lambda$ as $0.08 \pm 0.01$ and $\lambda\langle\omega^2\rangle$ as $29 \pm 4\ meV^2$ at $T_e = 590\ K$ by using the pump-probe approach. Even though the degree of electron excitation is unknown, all of these results [3, 13, 18] are in the ranges of $\lambda$ and $\lambda\langle\omega^2\rangle$ variations of present results in Figs. 2(c) and 2(d), which verifies the $T_e$ dependent $\lambda$ and $\lambda\langle\omega^2\rangle$ calculated in this letter.

After obtaining the full set of $f|_{T_e}$, $g(\varepsilon)|_{T_e}$, $g(\varepsilon_F)|_{T_e}$ and $\lambda\langle\omega^2\rangle|_{T_e}$, the $T_e$ dependent $G_{e-ph}$ are calculated from Eq. (7). As seen in Fig. 2(e), $G_{e-ph}$ (calculated in the present letter) firstly shows a steep increment to $5.6 \times 10^{18}\ W/(m^3K)$ at $T_e = 1.4 \times 10^4\ K$. The slight decrease of $G_{e-ph}$ when $1.4 \times 10^4\ K < T_e \leq 3.0 \times 10^4\ K$ and increase when $T_e > 3.0 \times 10^4\ K$ disagree with the profiles of $\lambda\langle\omega^2\rangle$ and $\lambda$ shown in Figs. 2(c) and 2(d). $\lambda\langle\omega^2\rangle|_{T_e=1.4\times10^4\ K}$ is 2.4 times of $\lambda\langle\omega^2\rangle|_{T_e=3.0\times10^4\ K}$, whereas, $G_{e-ph}|_{T_e=1.4\times10^4\ K}$ is only 1.5 times of $G_{e-ph}|_{3.0\times10^4\ K}$. Considering $g(\varepsilon_F)$ shown in Fig. 2(b), it can be concluded that the inverse of $g(\varepsilon_F)$ plays the dominating role of slowing down the decrement of $G_{e-ph}$ when $1.4 \times 10^4\ K < T_e \leq 3.0 \times 10^4\ K$. $g(\varepsilon_F)$ even facilitates the increment of $G_{e-ph}$ when $T_e > 3.0 \times 10^4\ K$. Compared with $G_{e-ph}$ computed from treating $g(\varepsilon)$ and $\lambda\langle\omega^2\rangle$ as constants by Lin and Zhigilei, and Chen's phenomenological estimation [5] of $G_{e-ph} = G_0[A_e/B_l(T_e + T_l) + 1]$ (where $G_0 = 1.0 \times 10^{17}\ W/(m^3K)$ [19], $A_e = 1.75 \times 10^7\ /(sK^2)$, $B_l = 1.98 \times 10^{11}\ /(sK)$ [20], and $T_l = $ $2835\ K$ (the boiling point of copper)), the $G_{e-ph}$ calculated in the present work shows great discrepancies. In Lin and Zhigilei's work [4], the effects of $\mu$ shift and $g$ variation are both eliminated in calculating $G_{e-ph}$, which are two crucial parameters contribute to the increase of $G_{e-ph}$ during electron excitation. In addition, the substitution of $\lambda\langle\omega^2\rangle = 29 \pm 4\ meV^2$ from experimental result in [3] is also another limitation hindering the comprehensive accounting of the electron-phonon interaction induced by electron excitation. Even though Chen *et al.* [5] took both the contributions of $T_e$ and $T_l$ to $G_{e-ph}$ into account by providing that $A_e(T_e + T_l) \ll B_l$, all the parameters $G_0$, $A_e$ and $B_l$ are obtained from the experiments [19, 20]. It should be noted that before vaporization, the actual $G_{e-ph}$ is lower than the result represented by the short dash line in Fig. 2(e). As explicitly seen in Fig. 2(e), when the electron temperature $T_e < 2 \times 10^3\ K$, the *ab initio* determined $G_{e-ph}$ is smaller than those treating $g(\varepsilon)$ and $\lambda\langle\omega^2\rangle$ as constants in calculating $G_{e-ph}$ and phenomenological $G_{e-ph}$. Both the *ab initio* determined $G_{e-ph}$ and constant $g(\varepsilon)$ and $\lambda\langle\omega^2\rangle$ calculated $G_{e-ph}$ show that the effective electron-phonon coupling factor does not change with $T_e$. Whereas, the phenomenological $G_{e-ph}$ presents monotonically increasing tendency with the increase of $T_e$. When $T_e > 1 \times 10^4\ K$, the *ab initio* determined $G_{e-ph}$ demonstrates almost 10 times of those treating $g(\varepsilon)$ and $\lambda\langle\omega^2\rangle$ as constants in calculating $G_{e-ph}$ and phenomenological $G_{e-ph}$.

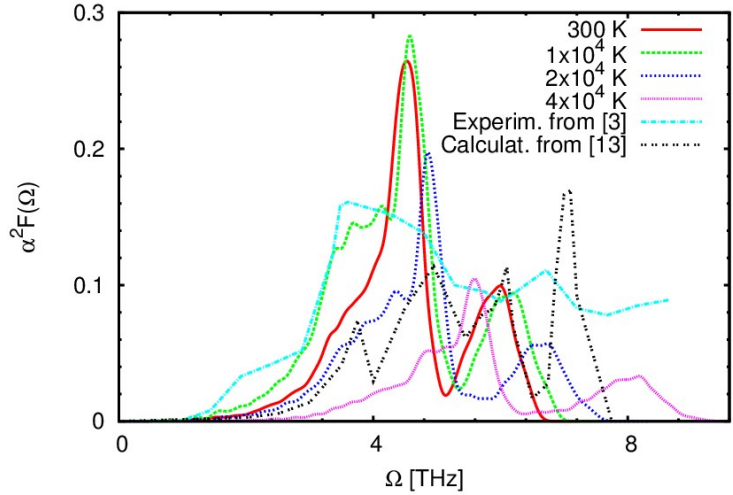

Fig. 3 Electron temperature $T_e$ dependent electron-phonon spectral function. The experimental result and ab initio calculated result are from [3] and [13], respectively.

In order to test the impact of pure *ab initio* determined $G_{e-ph}$ to the thermal energy transport from the electron subsystem to lattice subsystem, molecular dynamics (MD) and two-temperature model (TTM) coupled simulation is performed. The motion of copper atoms in the lattice subsystem is described by using MD, while the electron subsystem is characterized by a continuum energy equation in TTM. The embedded atom method (EAM) potential of copper [21] is chosen for its ability to represent the interatomic interaction. The lattice subsystem in MD and electron subsystem in TTM are coupled by $G_{e-ph}$. Detailed information of the MD-TTM coupled framework are seen elsewhere [22–25]. The laser pulse duration is set as $100\ fs$. The maximum intensity of laser pulse occurs at $t_0 = 15\ ps$. The absorbed laser energy is set at $0.3\ J/cm^2$. Optical penetration depth is chosen as $14.29\ nm$, for an incident laser with a laser wavelength of ~320 $nm$ [26]. The entire simulation lasts for 30 $ps$, which contained the first 5 $ps$ of room temperature control (in terms of canonical ensemble) to equilibrate the entire system and the second 5 $ps$ of micro-canonical ensemble verification to check the equilibration of lattice and electron subsystems. The entire system is constructed with 578.3840 $nm$, 3.6149 $nm$ and 3.6149 $nm$ in $x$-, $y$- and $z$- directions. Two empty spaces with thicknesses of 173.5152 $nm$ (occupying 30% of the entire length in $x$- direction) and 37.8384 $nm$ (occupying 10 % of the entire length in $x$- direction) are set to allow the film (thickness: 347.0304 $nm$) to expand during and after laser irradiation.

Figure 4(a) shows the spatial distribution of $T_e$ and $T_l$ from MD-TTM coupled simulation by employing $G_{e-ph}$ from *ab initio* calculation in the present letter. For comparison, Fig. 4(b) plots $T_e$ and $T_l$ calculated from MD-TTM coupled simulation by using the phenomenological $G_{e-ph}$ [5]. Both $T_l$ in the insets of Figs. 4(a) and 4(b) rise above the boiling point (2835 $K$) of copper. However, as seen from the starting points of $T_l$, no appreciable thermal expansions occur in either case. Due to larger $G_{e-ph}$ in Fig. 4(a) than that in Fig. 4(b), when the electrons are excited, larger amounts of thermal energy transfer from electron subsystem to lattice subsystem. This explains that why $T_l$ is greater in the inset of Fig. 4(a) than that in the inset of Fig. 4(b) from 15 $ps$ to 16 $ps$. As seen from $T_l$ at 18 $ps$, after laser heating, the time taking for lattice subsystem to get equilibration in Fig. 4(a) is much shorter than that in Fig. 4(b). At 30 $ps$, $T_l$ in the front side of the copper film is around 1,600 $K$ in Fig. 4(a). Whereas, for the case of Fig. 4(b), $T_l$ is around 1,200 $K$ in the corresponding location of the copper film. In addition, it can be seen the lattice heated region locates $x < 0.44$ in Fig. 4(a). However, larger lattice heated region ($x < 0.52$) is found in Fig. 4(b). As for $T_e$, it can be seen that $T_e$ in Fig. 4(a) presents smaller values than those in Fig. 4(b), which is mainly induced by greater $G_{e-ph}$ in the case of Fig. 4(a) during electron excitation.

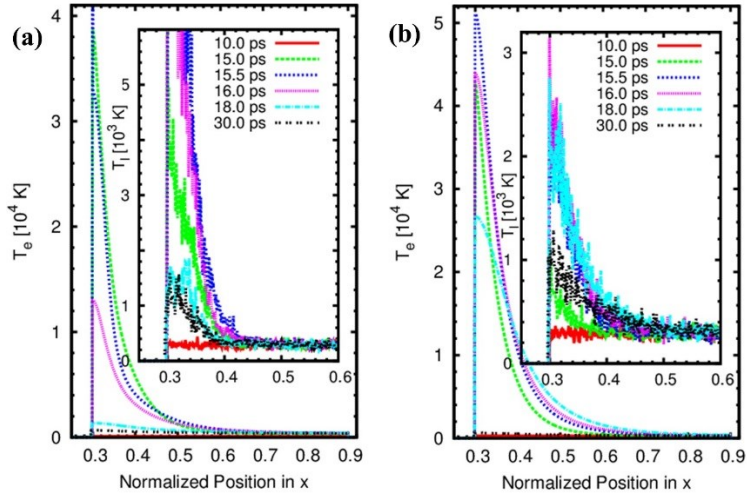

Fig. 4 Spatial distribution of electron temperature $T_e$ and lattice temperature $T_l$ from MD-TTM coupled simulation: (a) by using the electron-phonon coupling factor $G_{e-ph}$ calculated by pure ab initio calculation in the present work (b) by using Chen's phenomenological $G_{e-ph}$ [5].

## 4. Conclusions

In summary, it is concluded that both the previous treatment of $G_{e-ph}$ by using $T_e$ independent $g(\varepsilon)$ and $\lambda\langle\omega^2\rangle$ in Eq. (7) [4] and the phenomenological $G_{e-ph}$ [5] underestimate the effective electron-phonon interaction process. To take the variations of $g(\varepsilon)$, $f(\varepsilon)$ and $\alpha^2 F(\Omega)$ during electron excitation are essential. The MD-TTM coupled simulation shows the results by ultilizing pure *ab initio* calculated $G_{e-ph}$ giving a shallower heated region and faster energy transfer (as a result of lower $T_e$ and higher $T_l$) than those using phenomenological $G_{e-ph}$. Further experimental study is suggested to investigate and compare the results of laser heated copper film. The accurate determination of $G_{e-ph}$ in the presented letter leaves no empirical $G_{e-ph}$ in the MD-TTM coupled simulation and empowers the multi-scale modeling of laser material interaction.

## Acknowledgments

xSupport for this work by the U.S. National Science Foundation under grant number CBET-133611 is gratefully acknowledged.